\documentclass[prl,twocolumn,amsmath,amssymb,superscriptaddress]{revtex4}
\usepackage{graphicx}% Include figure files
\usepackage{bm,color}
\usepackage{amsmath,amssymb}

\begin{document}

\title{Blind quantum computation protocol in which Alice only
makes measurements} 

\author{Tomoyuki Morimae}
\affiliation
{ASRLD Unit, Gunma University, 
1-5-1 Tenjin-cho Kiryu-shi Gunma-ken, 376-0052, Japan}
\affiliation
{Department of Physics, 
Imperial College London, London SW7 2AZ, UK}
%\affiliation{
%IRCS,
%Tokyo Institute of Technology,
%2-12-1 Ookayama, Meguro-ku, Tokyo 152-8550, Japan
%}
\author{Keisuke Fujii}
\affiliation{
The Hakubi Center for Advanced Research, Kyoto University, 
Yoshida-Ushinomiya-cho, Sakyo-ku, Kyoto 606-8302, Japan}
\affiliation{
Graduate School of Informatics, Kyoto University, 
Yoshida Honmachi, Sakyo-ku, Kyoto 606-8501, Japan}
\affiliation{
Graduate School of Engineering Science, Osaka University, Toyonaka,
Osaka 560-8531, Japan
}
\date{\today}
            
\begin{abstract}
Blind quantum computation is a new secure quantum computing protocol which
enables Alice who does not have 
sufficient quantum technology to delegate her quantum computation
to Bob who has a fully-fledged quantum computer in such a way that
Bob cannot learn anything about Alice's input, output, and
algorithm.
In previous protocols, Alice needs to have a device
which generates quantum states, such as single-photon states.
Here we propose another type of blind computing protocol where
Alice does only measurements, such
as the polarization measurements with a threshold detector.
In several experimental setups, such as optical systems,
the measurement of a state is much easier than
the generation of a single-qubit state.
Therefore our protocols ease Alice's burden.
Furthermore, the security of our protocol
is based on the no-signaling principle, which is more
fundamental than quantum physics.
Finally, our protocols are device independent
in the sense that
Alice does not need to trust her measurement device
in order to guarantee the security.

\end{abstract}
\maketitle  
%------------------------------------

A first-generation quantum computer will 
be implemented in the ``cloud" style, 
since
only limited number of groups, such as governments and huge
industries, will be able to possess it.
How can a client
of such a cloud quantum computing
assure the security of his/her privacy?
Protocols of blind quantum 
computation~\cite{Childs,Arrighi,blindcluster,Aha,TVE,Vedran,Barz,
blind_Raussendorf,FK,blind_cv,Sueki} provide a solution. 
Blind quantum computation is a new secure quantum computing protocol
which enables a client
(Alice) who has only a classical computer or
a primitive quantum device which is not sufficient for universal
quantum computation
to delegate her computation to a server (Bob) 
who has a fully-fledged quantum computer without leaking
any Alice's privacy (i.e., which algorithm
Alice wants to run, which value Alice inputs, and what is the 
output of the computation) 
to Bob~\cite{Childs,Arrighi,blindcluster,Aha,TVE,Vedran,Barz,blind_Raussendorf,FK,blind_cv,Sueki}.

The first example of blind quantum computation was 
proposed by Childs~\cite{Childs} where the quantum circuit model was used,
and the register state was encrypted with quantum
one-time pad~\cite{onetimepad} so that Bob who performs
quantum gates learns nothing about information in the quantum register.
In this method, however, Alice needs to have a quantum memory and 
the ability to perform the SWAP gate. 
The protocol proposed by Arrighi and Salvail \cite{Arrighi} is that
for the calculation of certain classical functions, i.e., not the 
universal quantum computation, and it requires 
Alice to prepare and measure
multi-qubit entangled states.
Furthermore, it is cheat-sensitive, i.e., Bob can gain information
if he does not mind being caught. 
Finally, in their protocol, Bob knows the unitary which Alice wants
to implement.
Aharonov, Ben-Or and Eban's protocol \cite{Aha} requires a constant-sized quantum computer with a quantum memory for Alice.

On the other hand, in 2009,
Broadbent, Fitzsimons and Kashefi~\cite{blindcluster}
proposed a new blind quantum computation protocol which uses
the one-way model~\cite{cluster,cluster2,Raussendorf_PhD,hammer}.
In their protocol, all Alice needs 
are a classical computer and a primitive quantum device, which emits
randomly rotated single-qubit states. In particular, Alice does not require any
quantum memory and the protocol is unconditionally secure
(i.e., Alice's input, output, and algorithm are secret to Bob
whatever Bob does). 
Recently, this protocol has been experimentally demonstrated
in an optical system~\cite{Barz}.
Furthermore, this innovative protocol has inspired several new other
protocols which can enjoy more robust blind quantum computation.
In Ref.~\cite{TVE}, two protocols
which enable blind measurement-based quantum
computation on the Affleck-Kennedy-Lieb-Tasaki (AKLT) 
state~\cite{AKLT,MiyakeAKLT}
have been proposed. In Ref.~\cite{blind_Raussendorf},
a protocol of the blind topological measurement-based quantum 
computation~\cite{Raussendorf_PRL,Raussendorf_NJP,Raussendorf_Ann}
has been proposed.
Due to the topological protection, it is fault-tolerant~\cite{Raussendorf_PRL,Raussendorf_NJP,Raussendorf_Ann}. 
The error threshold of the blind topological model
has been shown 
to be comparable to that of the original~\cite{Raussendorf_NJP,Raussendorf_PRL} 
(i.e., non-blind) topological quantum
computation~\cite{blind_Raussendorf}.

Before starting the main part of this paper,
let us quickly review the
protocol of Ref.~\cite{blindcluster}. In this protocol,
Alice and Bob share a classical channel and a quantum channel.
The protocol runs
as follows: (1) Alice prepares
randomly-rotated single-qubit states
$\{|\theta_j\rangle\equiv
|0\rangle+e^{i\theta_j}|1\rangle\}_{j=1}^N$,
where $\theta_j\in{\mathcal A}\equiv\{\frac{k\pi}{4}|k=0,1,...,7\}$
is a random angle,
and sends them to Bob through the quantum channel.
(2) Bob creates a certain  
two-dimensional graph state, which is called
the brickwork state~\cite{blindcluster}, 
by applying the C$Z$ gates among
$\{|\theta_j\rangle\}_{j=1}^N$.
(3) Alice calculates 
the measurement angle
on her classical computer,
and
sends it to Bob through the classical channel.
(4) Bob performs the measurement in that angle,
and returns the measurement result to Alice.
(5) They repeat (3)-(4) until 
the computation is finished.

If Bob is honest, Alice obtains the correct answer
of her desired quantum computation~\cite{blindcluster}.
%If Bob is not cooperative, the correctness of the final output
%is no longer guaranteed. However,
Furthermore, it was shown that whatever evil Bob does, Bob cannot learn anything
about Alice's input, output, and algorithm~\cite{blindcluster}.
%(An intuitive understanding of the security of this protocol is as follows.
%Let us assume that Alice wants Bob to measure a particle of
%the graph state in the angle $\phi$.
%If Alice directly sends $\phi$ to Bob, Bob knows $\phi$, and
%therefore Bob can gain some information about Alice's algorithm.
%Therefore, Alice sends $\delta=\phi+\theta+r\pi$ to Bob,
%where $r\in\{0,1\}$ is a random number which comes from a certain
%technical reason which we do not explain here~\cite{blindcluster}.
%Since the particle to be measured is pre-rotated
%by $\theta$ in the above process (1),
%the particle is effectively measured in the angle $\phi+r\pi$,
%when Bob does the measurement in the angle $\delta$.
%The effect of $r\pi$ is just the flip of the measurement result.
%In this way, Alice can have Bob do the measurement in the angle $\phi$
%without allowing Bob to know the value of $\phi$.) 

The motivation of the blind quantum computation is to enable
Alice, who does not have any sophisticated technology
and enough knowledge,
to perform universal quantum computation.
Therefore, there are two important goals. One is
to make Alice's device
as classical as possible, 
since Alice is not expected to have any expensive laboratory
which can maintain the coherence of complicated quantum experimental setups.
The other is to exempt Alice
from the precise verification of her device,
since Alice is not expected to have enough technology and knowledge
to verify her device. Such a verification is important
since she might buy the device from a
company which is under the control of Bob, and therefore the device might
not work as Alice expects. 
For example, if Alice is supposed to send a single-photon to Bob,
Alice must confirm that more than two identical photons are not sent to 
Bob,
since otherwise
Bob might be able to gain some information by using, e.g., 
the photon-number-splitting (PNS) attack~\cite{PNS0,PNS1,PNS2,PNS3},
which is a well-known technique in quantum key distribution (QKD).
In Ref.~\cite{Vedran}, a first step to the first goal, namely
making Alice's device more
classical, 
was achieved.
They proposed an ingenious protocol of the blind quantum computation
in which what Alice needs to prepare
are not single-photon states but coherent states. Since
coherent states are considered
to be
more classical than single-photon states,
their protocol allows Alice's device to be more classical.
%However, in their protocol, Bob is required to perform
%the non-demolition photon-number measurement, which is not easy
%with the current technology.

In this paper, 
we show that Alice who has only a measurement device
can perform the blind quantum computation. 
In several experimental setups, such as quantum optical systems, 
the measurement of a state, e.g., the polarization measurement of photons
with a threshold detector,
is much easier than the generation of a single-qubit state,
such as a single-photon state.
Therefore, our results achieve the above mentioned first goal, namely
making Alice's device more classical.
As we will see later,
our protocols can cope with the particle loss in the quantum channel
between Alice and Bob
and the measurement inefficiencies of Alice's measurement device.
It also means that our protocols allow Alice's device to be more classical.

We propose two new protocols, Protocol 1 and Protocol 2.
Importantly, the security of Protocol 1
is based on the no-signaling principle~\cite{no-signaling},
which is more fundamental than quantum physics~\cite{no-signaling}.
Therefore, even if Bob does a super-quantum (but no-signaling) 
attack, Alice's privacy
is still guaranteed.
Furthermore, the device-independence~\cite{device-independent}
is attained for the security of Protocol 1.
Hence the above second goal is also achieved. 
The security based on the no-signaling principle 
and device-independence are important subjects in
quantum key distribution, and much researches have been done
within the decade~\cite{BHK}.
However, Protocol 1 cannot cope with a high channel loss.
Protocol 2 can tolerate any high channel loss,
but the device-independence becomes weaker.

%----------------------
{\it Protocol 1}.---
Our first protocol runs as follows:
(1) Bob prepares a resource state of measurement-based
quantum computation. Any resource state can be used for this purpose.
For example,
the two-dimensional cluster state~\cite{cluster,cluster2,Raussendorf_PhD}, 
the three-dimensional cluster state
for the topological quantum computation~\cite{Raussendorf_PRL,Raussendorf_NJP,Raussendorf_Ann}, 
the thermal equilibrium states of a nearest-neighbour two-body
Hamiltonian with spin-2 and spin-3/2 
particles~\cite{Li_finiteT} 
or solely with spin-3/2 particles~\cite{FM_finiteT} at a finite temperature
for the topological measurement-based quantum computation,
resource states for the quantum computational tensor 
network~\cite{Gross1,Gross2,Gross3,upload,stringnet,FM_correlation,CPTP},
the one-dimensional or two-dimensional
AKLT states~\cite{MiyakeAKLT,MiyakeAKLT2d,WeiAKLT2d}, 
the tri-cluster state~\cite{tricluster},
and
states in the Haldane phase~\cite{Miyake_Haldane}.
%(2) Bob lets Alice know which resource state he has.
(2) Bob sends a particle of the resource 
state to Alice through the quantum channel.
(3) Alice measures the particle in a certain angle
which is determined by the algorithm in
her mind.
They repeat (2)-(3) until 
the computation is finished.

Obviously, at the end of the computation,
Alice obtains the correct answer of
her desired quantum computation if Bob is honest,
since what Alice and Bob did is nothing but a usual measurement-based 
quantum computation. 
(It is something like the following story:
Alice and Bob are in the same laboratory.
The preparation and the maintenance of the resource state, which
are boring routines,
are done by a student Bob, whereas the most exciting
part of the measurement-based quantum computation, 
namely the measurements and the collection of data are done by his boss Alice.
Somehow, there is no communication between the boss and
the student.)

It is also easy to understand that whichever states evil Bob prepares
instead of the correct resource state,
and whichever states evil Bob sends to Alice,
Bob cannot learn anything about Alice's information,
since Alice does not send any signal to Bob
and therefore because of the no-signaling principle~\cite{no-signaling}
Bob cannot gain any information about Alice
by measuring his system~\cite{teleportation}:
If Alice could transmit some information to Bob by measuring
her system, it contradicts to the no-signaling principle~\cite{noQC}.
(Note that we assume 
there is no unwanted leakage of information from Alice's laboratory.
For example, Bob cannot bug Alice's laboratory. 
It is the standard assumption in the quantum key distribution~\cite{QKD}.)
In Sec.~I of Appendix, we give the mathematical proof of the security
of Protocol 1 based on the no-signaling principle.

Protocol 1 has four advantages.
First, unlike Ref.~\cite{blindcluster},
no random-number generator is required for Alice.
This is advantageous since it is not easy to generate
completely random numbers, and the random-number generator might
be provided by a company under the control of Bob.
Second,
the security of the protocol is device independent
in the sense that Alice does not need to trust her measurement device
in order to guarantee the security,
since whatever Alice's device does, 
Bob cannot gain any information about Alice's
computation (due to the no-signaling principle)
as long as there is no unwanted leakage of information from Alice's laboratory,
which is the standard assumption in quantum key distribution~\cite{QKD}.
Third, the proof of the security is intuitive and very simple,
and 
it is based on the no-signaling principle~\cite{no-signaling},
which is more fundamental than quantum physics~\cite{no-signaling}.
(Even if quantum physics is violated in a future, 
Protocol 1 survives
as long as the no-signaling principle holds.)
Finally, any model of measurement-based quantum computation
(such as the cluster model~\cite{cluster,cluster2,Raussendorf_PhD}, 
the AKLT models~\cite{MiyakeAKLT,MiyakeAKLT2d,WeiAKLT2d}, 
and the topological model~\cite{Raussendorf_PRL,Raussendorf_NJP,Raussendorf_Ann}, etc.) 
can be directly changed into a blind model:
Bob has only to let Alice do measurements.
(On the other hand, in Ref.~\cite{TVE}, 
many complicated procedures are
required to make
the AKLT measurement-based quantum computation blind.)
Since no modification
is required to make a model blind, the advantage of a model
is preserved when it is changed into a blind model.
For example, an advantage of doing the measurement-based quantum
computation on the AKLT states is that the quantum computation is protected
by the energy gap of a physically natural Hamiltonian~\cite{MiyakeAKLT,MiyakeAKLT2d,WeiAKLT2d}.
If the AKLT model is used in Protocol 1, 
Bob who prepares and maintains the resource AKLT
state can enjoy that advantage, i.e., Bob's state is protected
by the energy gap.
This is also the case for the models of Refs.~\cite{Li_finiteT,FM_finiteT}:
If these models are used in Protocol 1,
Bob can enjoy the advantage of these models, i.e., 
Bob does not need to keep his state in the ground state;
His state is allowed to be the equilibrium state at a finite temperature.
%However, in this case, Bob has just only to do the usual AKLT measurement-based quantum computation
%except for the measurements which are done by Alice.
%Therefore, the resource state is protected by the energy gap of the AKLT Hamiltonian which Bob has.
%Furthermore, it was shown there that the measurement-based quantum computation
%cannot enjoy the energy-gap protection in the single-sever scheme.
%However, in this protocol,
%Bob can enjoy the energy gap protection; Bob has only to keep
%his state in the ground state of the AKLT Hamiltonian.
%Moreover, the model of the topological measurement-based quantum computation
%on a finite-temperature equilibrium state can be directly
%done in our protocol; Bob has only to keep his system sufficiently
%cold.

A disadvantage of Protocol 1 is that 
the quantum channel between Alice and Bob must not be too lossy.
(Throughout this paper, ``the channel loss" includes
the detection inefficiency of Alice's device,
since the detection inefficiency behaves like the channel loss.)
On the other hand, in the previous protocols~\cite{blindcluster,TVE,Barz,blind_Raussendorf,FK} where Alice sends randomly rotated particles to Bob,
the high loss rate of
the quantum channel is not crucial, since if Bob does not
receive a particle due to the 
loss in the quantum channel, Bob has only to ask
Alice 
to again generate and send another state with another random angle.
One way of overcoming that disadvantage
of Protocol 1
is
to use a model which can cope with the particle loss.
For example, it was shown in Ref.~\cite{Barrett_loss} that the topological measurement-based quantum computation~\cite{Raussendorf_PRL,Raussendorf_NJP,Raussendorf_Ann}
can cope
with the heralded particle loss if the loss probability
is below the threshold.
If Bob uses this model,
Alice and Bob can perform Protocol 1 
without suffering from the particle loss
as long as
the loss rate of the quantum channel between
Alice and Bob (and that of Alice's device) is below  
the loss threshold calculated in Ref.~\cite{Barrett_loss}.

%------------------
{\it Protocol 2}.---
If we want to have a protocol which is tolerant against
any high channel loss rate, we have to give up the perfect
no-singling, since Alice has to send some message to Bob
when a particle is lost. 

One way of making Protocol 1 tolerant against any high channel
loss is to use the quantum teleportation.
Let us consider the following protocol:
(1) Bob prepares a resource state.
(2) He creates a Bell pair, and sends a half of it to Alice.
(3) If the particle is lost, Alice asks Bob to send it again.
If Alice receives the particle, Alice lets Bob know it.
(4) Bob teleports a particle of his resource state to Alice by
using the Bell pair.
(5) Bob sends the measurement result of the teleportation to Alice.
(6) Alice measures the teleported particle in an angle which is determined
by her algorithm (and Bob's teleportation result).

This protocol is a modified version of Protocol 1
where Bob teleports a particle of his resource state
instead of directly sending it to Alice.
This protocol is loss tolerant, since if a half of a Bell pair is lost in 
the channel, Bob has only to send it again.
However, this protocol has a huge problem:
Alice has to have a single-particle quantum memory,
since Alice's measurement must be done after
Bob's teleportation (otherwise Alice cannot correct byproducts
created by Bob's teleportation). 
Such a quantum memory does not need to have a long coherence
time since the quantum teleportation can be done quickly,
but still the requirement of a quantum memory is disadvantageous to Alice.

Here, we introduce Protocol 2, which can avoid such a quantum memory.
This is our second main result of this paper.
The basic idea of Protocol 2 is that Alice ``prepares" rotated states
which ``encode" algorithm
in Bob's place, and Bob performs a layer-by-layer
measurement-based quantum computation with these rotated states.
Protocol 2 runs as follows:
(1) Bob creates a Bell pair, and sends a half of it to Alice
through the quantum channel. 
(2) If Alice does not receive it, because of the channel loss,
Alice asks Bob to send it again and goes back to (1).
(3) If Alice receives the particle, she measures it in
the basis $\{|0\rangle\pm e^{-i\theta}|1\rangle\}$,
where $\theta$ is a certain angle (not a random angle) determined
by the algorithm which Alice wants to run.
($\theta=0,\pi/2$ for Clifford gates, and $\theta=\pi/4$ for a non-Clifford
gate. Details will be explained in Sec.~II of Appendix.)
In this way, Alice can ``prepare" a state which encodes the angle
of the algorithm in Bob's place.
(4) Bob couples the half of the Bell pair which he has to a qubit
of his register state
by using the $CZ$ gate, and 
measures the qubit in the register state
in the $\{|+\rangle,|-\rangle\}$ basis.
This $X$-basis measurement implements the quantum gate.
(5) Bob sends the result of this $X$-basis measurement
to Alice through the classical channel.
(6) They repeat (1)-(5) until the computation is finished.

In Sec.~III of Appendix, 
we show the blindness of Protocol 2:
whichever states evil Bob prepares
and whichever states evil Bob sends to Alice, Bob
cannot learn anything about Alice's information. 

One might think Alice's measurement in step (3) has to be delayed
until the end of step (5) so that she can feed-forward Bob's measurement
outcome as usual one-way model.
However, this is not the case.
In Protocol 2, by properly choosing the measurement basis,
we give a way to postpone Alice's feed-forwarding until her
subsequent measurements, and hence she does not have to wait for Bob's
measurement outcome. This means that no quantum memory is required
for Alice.
In Sec.~II of Appendix,
we give a detailed explanations about how Alice should measure
particles. 

In this way, we can obtain a protocol which is loss tolerant.
However, as we have mentioned earlier, there is a trade-off between
loss tolerance and no-signaling.
Protocol 2 is no longer no-signaling.
In Protocol 1, no signal is transmitted from Alice to Bob,
and therefore the no-signaling is completely satisfied.
However, in Protocol 2, the message whether Alice receives a particle
or not
is sent from Alice to Bob,
and therefore the no-signaling is no longer satisfied.
One might think that such a message is not directly related to
Alice's measurement angles, and therefore the situation is
``quasi" no-signaling.
However, if we want to show the device-independence security, 
a special care is necessary.
In Ref.~\cite{prisoner} it was shown that in quantum key distribution
if Alice and Bob
use the same measuring device many times, some secret information can be
broadcasted through the ``legal" channel.
Similar attack can be considered in our Protocol 2. For example, let us
assume that Alice does the measurement in the angle $5\pi/4$.
Then, the measuring device remembers the number 5, and 
pretends to loss the particle fifth times.
Then Alice sends the message, ``the particle is lost", 
to Bob fifth times, and from that fact,
Bob can know the number 5.
One way of avoiding such an attack is, as is explained in Ref.~\cite{prisoner},
to discard the measuring device after using it, and
to use new device for every measurement.
The other way, which can be used in our Protocol 2, is
that Alice generates a random bit $b$ when Bob sends a particle, and behaves
as if the particle is lost (arrived) when $b=0$ (1).
In this case, Alice needs a random number generator,
but the evil measuring device can no longer do that attack. 

{\it Discussion}.---
In this paper, we have proposed protocols of blind
quantum computation for Alice who does
only measurements,
such as the polarization measurement with a threshold detector.
In quantum optics, for example, the state measurement is 
much easier than the single-qubit state
generation. Therefore our scheme makes Alice more classical
than the previous protocols~\cite{blindcluster,TVE,blind_Raussendorf}
in certain experimental setups, such as optical systems.
In the protocol of Ref.~\cite{Vedran}, Bob is required to perform
the non-demolition photon-number measurement, which is not easy
with the current technology.
In our protocols, on the other hand,
Bob is not required to have such an additional high technology.

We have proposed two protocols, Protocol 1 and Protocol 2.
Procotol 1 is simple, its security is based on the no-signaling principle,
and it satisfies the device-independent security.
However, it can not tolerate a high channel loss rate.
On the other hand, Protocol 2 can tolerate any high channel loss rate,
although it is more complicated than Protocol 1 and no longer
no-signaling.
Appropriate one should be chosen depending on the situation.

Finally, let us briefly discuss about the 
verification~\cite{Aha,blindcluster,FK,topoveri}
of blind quantum computing.
The verification is a way of Alice checking whether Bob
is honestly following her protocol~\cite{Aha,blindcluster,FK,topoveri}.
It is important for blind quantum computation, since
evil Bob can just destroy the computation and Alice might
accept wrong computational results.
Methods of the verification for blind quantum computation
of Ref.~\cite{blindcluster} 
were already proposed in Refs.~\cite{blindcluster,FK}. 
Can we do the verification for our measuring Alice blind
quantum computation?
One simple way of doing verification is that
Alice randomly chooses some subsystem of the resource state
and measures the stabilizer operators
in order to check whether Bob correctly creates the resource graph state.
Recently, more efficient way of doing verification for
our measuring Alice protocol has been proposed in
Ref.~\cite{topoveri}, 
which uses the previous verification methods of
Aharonov, Ben-Or, and Eban~\cite{Aha}
and Fitzsimons and Kashefi~\cite{FK}.
By using that verification method, Alice can check whether
Bob is honestly doing computation or not.

TM was supported by JSPS.
KF was supported by MEXT Grant-in-Aid for Scientific 
Research on Innovative Areas 20104003. 
We acknowledge Matty J. Hoban for
bringing our attention to Ref.~\cite{prisoner}.

\begin{widetext}

\section{Blindness of Protocol 1}
We assume that the initial state of the computation is the
standard state $|0...0\rangle$, and the preparation of the
input state is included in the computational part.
Therefore, we can assume without loss of generality
that what Alice wants to hide are the computation angles
of the measurement-based quantum computation
and the final output of the computation.
Intuitively, a protocol is blind if Bob, given all the classical and quantum information during the protocol, 
cannot learn anything about Alice's computational angles and the
output~\cite{blindcluster,TVE,Vedran,blind_Raussendorf}.

{\bf Definition}:
{\it
In this paper,
we call
a protocol is blind if 
\begin{itemize}
\item[(B1)]
The conditional probability distribution of  
Alice's computational angles, 
given all the classical information Bob can obtain during the protocol, 
and given the measurement results of any POVMs which Bob may perform
on his system at any stage of the protocol,
is equal to the a priori probability distribution of
Alice's computational angles,
and
\item[(B2)]
%The final classical output is one-time padded to Bob.
The conditional probability distribution of  
the final output of Alice's algorithm, 
given all the classical information Bob can obtain during the protocol, 
and given the measurement results of any POVMs which Bob may perform
on his system at any stage of the protocol,
is equal to the a priori probability distribution of
the final output of Alice's algorithm.
\end{itemize}
}
$\blacksquare$
Intuitively, this means that Bob's ``certainty" about Alice's information
does not changed even if Bob does POVM on his system.

{\bf Theorem 1}: 
{\it Protocol 1 satisfies (B1).}

{\bf Proof}: 
Let $A$ be the random variable which represents Alice's measurement angles,
and $B$ be the random variable which represents the
type of the POVM which Bob performs on his system.
%Let $M_A$ be the random variable which represents Alice's measurement results
Let $M_B$ be the random variable which represents 
the result of Bob's POVM.
The two-party system is called no-signaling~\cite{no-signaling} 
from Alice to Bob iff
%\begin{eqnarray*}
%\sum_{m_1}P(M_A=m_A,M_B=m_B|U_1=u_1,U_2=u_2)
%&=&\sum_{m_1}P(M_A=m_A,M_B=m_B|U_1=u'_1,U_2=u_2),\\
%\sum_{m_2}P(M_A=m_A,M_B=m_B|U_1=u_1,U_2=u_2)
%&=&\sum_{m_2}P(M_A=m_A,M_B=m_B|U_1=u_1,U_2=u'_2).
%\end{eqnarray*}
\begin{eqnarray*}
P(M_B=m_B|A=a,B=b)
&=&P(M_B=m_B|A=a',B=b),
%P(M_A=m_A|A=a,B=b)
%&=&P(M_A=m_A|A=a,B=b'),
\end{eqnarray*}
for all $m_B$, $a$, $a'$, and $b$.
Then,
\begin{eqnarray*}
P(A=a|B=b,M_B=m_B)
&=&
\frac{
P(M_B=m_B|A=a,B=b)
P(A=a,B=b)
}{P(B=b,M_B=m_B)}\\
&=&
\frac{
P(M_B=m_B|A=a,B=b)
P(A=a|B=b)P(B=b)
}{P(B=b,M_B=m_B)}\\
&=&
\frac{
P(M_B=m_B|A=a',B=b)
P(A=a'|B=b)P(B=b)
}{P(B=b,M_B=m_B)}\\
&=&
P(A=a'|B=b,M_B=m_B).
\end{eqnarray*}
This means that Bob cannot learn anything about Alice's measurement angles.
$\blacksquare$

{\bf Theorem 2}:
{\it Protocol 1 satisfies (B2).}

{\bf Proof}:
Let $O$ be the random variable which represents the output of Alice's 
algorithm,
and $B$ be the random variable which represents the
type of the POVM which Bob performs on his system.
Let 
$M_B$ be the random variable which represents 
the result of Bob's POVM.
Alice can change the output of her algorithm by
changing the input.
(For example, since what is implemented in the quantum computation
is a unitary operation, two input states which are orthogonal with each
other become two mutually-orthogonal output states.)
Because of the no-signaling principle,
\begin{eqnarray*}
P(M_B=m_B|O=o,B=b)=P(M_B=m_B|O=o',B=b)
\end{eqnarray*}
for all $m_B$, $o$, $o'$, and $b$.
Then,
\begin{eqnarray*}
P(O=o|B=b,M_B=m_B)
&=&
\frac{
P(M_B=m_B|O=o,B=b)
P(O=o,B=b)
}{P(B=b,M_B=m_B)}\\
&=&
\frac{
P(M_B=m_B|O=o,B=b)
P(O=o|B=b)P(B=b)
}{P(B=b,M_B=m_B)}\\
&=&
\frac{
P(M_B=m_B|O=o',B=b)
P(O=o'|B=b)P(B=b)
}{P(B=b,M_B=m_B)}\\
&=&
P(O=o'|B=b,M_B=m_B).
\end{eqnarray*}
Therefore, Bob cannot learn anything about the output
of Alice's algorithm.
$\blacksquare$

\section{Correctness of Protocol 2}

%The set
%$\{H,S\equiv R_{\pi/2},CNOT\}$
%generates the Clifford group.
%Furthermore, the Clifford group plus $T\equiv R_{-\pi/4}$
%realize the universal quantum computation.

Protocol 2 runs as follows:
(1) Bob prepares the Bell pair and sends the half of it to Alice.
(2) If Alice does not receive it, she asks Bob to try again,
and goes back to (1).
(3) If Alice receives the particle, she does the measurement in the 
\begin{eqnarray*}
\Big\{\frac{1}{\sqrt{2}}\big(
|0\rangle\pm e^{-i\theta}|1\rangle
\big)\Big\}
\end{eqnarray*}
basis.
How to choose $\theta$ will be explained later.
After her measurement, Bob has the state
\begin{eqnarray*}
Z^aR_\theta|+\rangle,
\end{eqnarray*}
where $|\pm\rangle=\frac{1}{\sqrt{2}}(|0\rangle\pm|1\rangle)$,
$a\in\{0,1\}$ is Alice's measurement result, and
\begin{eqnarray*}
R_\theta=\Big(
\begin{array}{cc}
1&0\\
0&e^{i\theta}
\end{array}
\Big).
\end{eqnarray*}
Note that $R_\theta X=e^{i\theta}XR_{-\theta}$.
(4) Bob 
creates the state
\begin{eqnarray*}
CZ_{1,2}\Big(|\psi\rangle_1\otimes Z^aR_\theta|+\rangle_2\Big),
\end{eqnarray*}
where
$CZ_{1,2}$ is the C$Z$ gate between the first
and the second qubits and $|\psi\rangle$ is any Bob's state.
(5) Bob
performs the measurement in the basis $\{|+\rangle,|-\rangle\}$
on the first qubit.
Since $Z^aR_\theta$ commutes with $CZ_{1,2}$,
Bob obtains
\begin{eqnarray}
Z^aR_\theta X^mH|\psi\rangle_2
\label{Bobobtains}
\end{eqnarray}
if the measurement result is $m\in\{0,1\}$,
where $H$ is the Hadamard gate.

For $\theta=0$,
Eq.~(\ref{Bobobtains}) becomes
\begin{eqnarray*}
Z^aX^mH|\psi\rangle_2.
\end{eqnarray*}

For $\theta=\pi/2$,
Eq.~(\ref{Bobobtains}) becomes
\begin{eqnarray*}
Z^aR_{\pi/2}X^mH|\psi\rangle_2
&=&Z^aX^mR_{(-1)^m\pi/2}H|\psi\rangle_2\\
&=&Z^aX^mZ^mR_{\pi/2}H|\psi\rangle_2\\
&=&Z^{a+m}X^mSH|\psi\rangle_2,
\end{eqnarray*}
where
$S=R_{\pi/2}$.

For $\theta=-\pi/4$,
Eq.~(\ref{Bobobtains}) becomes
\begin{eqnarray*}
Z^aR_{-\pi/4}X^mH|\psi\rangle_2
&=&Z^aX^mR_{-(-1)^m\pi/4}H|\psi\rangle_2\\
&=&
\left\{
\begin{array}{cc}
Z^aX^mTH|\psi\rangle_2&(m=0)\\
Z^aX^mT^\dagger H|\psi\rangle_2&(m=1),
\end{array}
\right.
\end{eqnarray*}
where
$T=R_{-\pi/4}$.

\if0
Note that
\begin{eqnarray*}
(PH)(PH)(PH)&=&PH,\\
(PH)(PH)(PSH)&=&PSH=PZS^\dagger H,\\
(PH)(PH)(PTH)&=&PTH,\\
(PSH)(PH)(PT^\dagger H)&=&PTH,\\
\end{eqnarray*}
where $P$ is a Pauli byproduct.
(Be careful that different Pauli byproducts are represented
by the same character $P$ for simplicity.)
Therefore, if Alice and Bob repeat the above (1)-(5)
three times,
$\{H,SH,S^\dagger H, TH\}$ can be deterministically implemented 
(up to some Pauli
byproduct)
on any state $|\psi\rangle$.

As is shown in Ref.~\cite{blindcluster},
universal quantum computation is possible on the brickwork
state~\cite{blindcluster}
if $\{H,HS,HS^\dagger,HT\}$ is possible.
Note that
\begin{eqnarray*}
(PH)(PH)(PH)(PH)(PH)&=&PH,\\
(PH)(PH)(PH)(PSH)(PH)&=&PHS,\\
(PH)(PH)(PH)(PS^\dagger H)(PH)&=&PHS^\dagger.
\end{eqnarray*}
Furthermore,
if $P'=X$ or $I$
\begin{eqnarray*}
(PH)(PH)(PH)(PTH)(P'H)&=&PHT
\end{eqnarray*}
and
if $P'=Z$ or $XZ$
\begin{eqnarray*}
(PH)(PSH)(PH)(PTH)(P'H)&=&
(PH)(PSH)(PH)(PT^\dagger)\\
&=&PHST^\dagger\\
&=&PHT.
\end{eqnarray*}
Therefore, if Alice and Bob repeat the above (1)-(5) for a constant
number of times,
they can perform the universal quantum computation.
\fi

Note that
\begin{eqnarray*}
(PH)(PH)(PH)&=&PH,\\
(PH)(PH)(PSH)&=&PSH,\\
(PH)(PSH)(PSH)&=&
PHSHSH=PZS\\
(PH)(PH)(PTH)&=&PTH,\\
(PSH)(PH)(PTH)&=&PT^\dagger H,\\
(PSH)(PH)(PT^\dagger H)&=&PTH,\\
(PH)(PH)(PT^\dagger H)&=&PT^\dagger H,
\end{eqnarray*}
where $P$ is a Pauli byproduct.
(Be careful that different Pauli byproducts are represented
by the same character $P$ for simplicity.)
This means that
the operations
\begin{eqnarray*}
\Big\{H,TH,T^\dagger H,SH,S\Big\}
\end{eqnarray*}
can be done deterministically (up to Pauli byproducts) if Alice and Bob repeat 
the above (1)-(5) three times.
Therefore,
if we consider the unit cell (Fig.~\ref{unitcell}),
the operations
\begin{eqnarray*}
\Big\{
I\otimes I,
SH\otimes I,
STH\otimes I,
ST^\dagger H\otimes I,
H\otimes I,
(CZ)(CNOT)
\Big\}
\end{eqnarray*}
can be implemented deterministically up to some Pauli byproducts
as is shown in Fig.~\ref{unitcell2}.
Note that this set is universal set, since
\begin{eqnarray*}
(PSH)(PH)&=&PS,\\
(PS)(PSTH)(P'H)&=&PT,\\
(PS)(PST^\dagger H)(P''H)&=&PT,\\
\end{eqnarray*}
where $P'$ is $I$ or $X$,
and
$P''$ is $Z$ or $XZ$.
As is shown in Fig.~\ref{tiling},
the unit cell can be tiled to create the universal
two-dimensional graph state
which resembles the brickwork state~\cite{blindcluster}.

\begin{figure}[htbp]
\begin{center}
\includegraphics[width=0.25\textwidth]{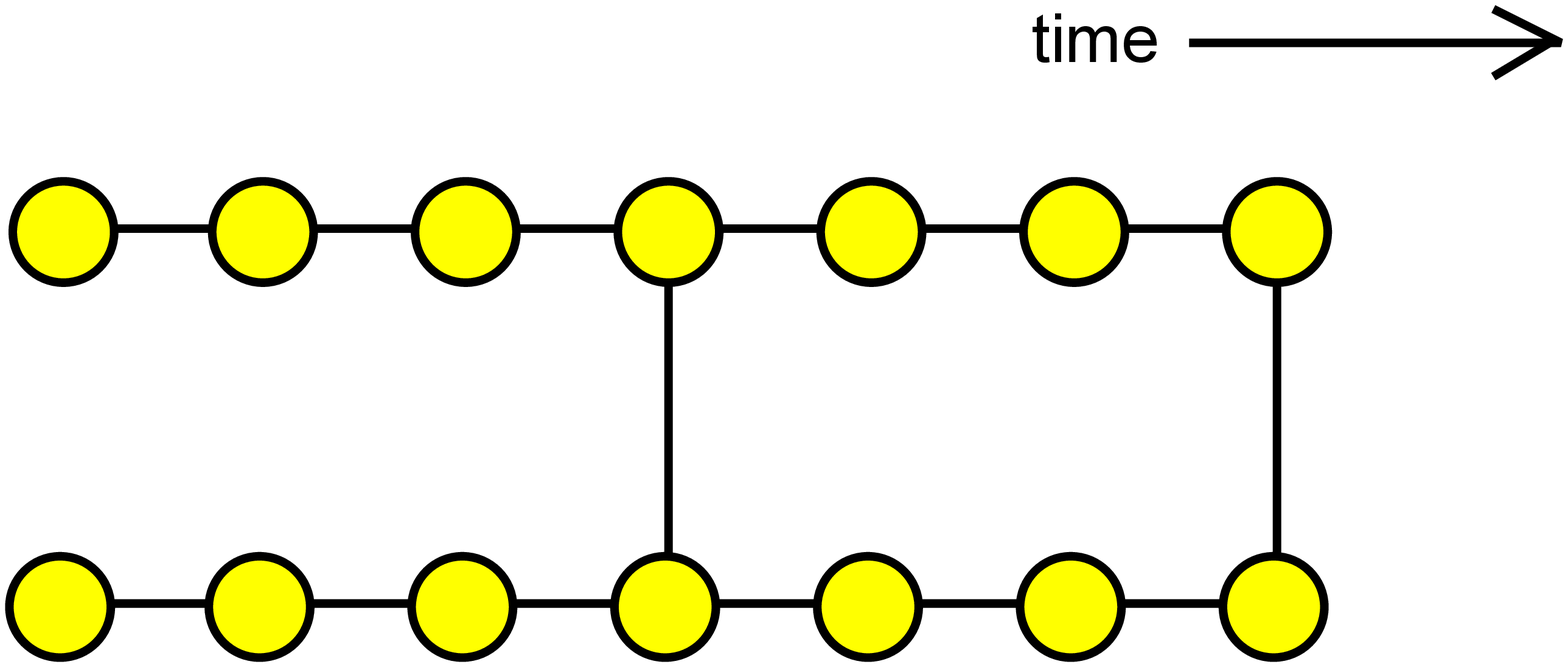}
\end{center}
\caption{(Color online.) 
The unit cell for Protocol 2.
} 
\label{unitcell}
\end{figure}

\begin{figure}[htbp]
\begin{center}
\includegraphics[width=0.6\textwidth]{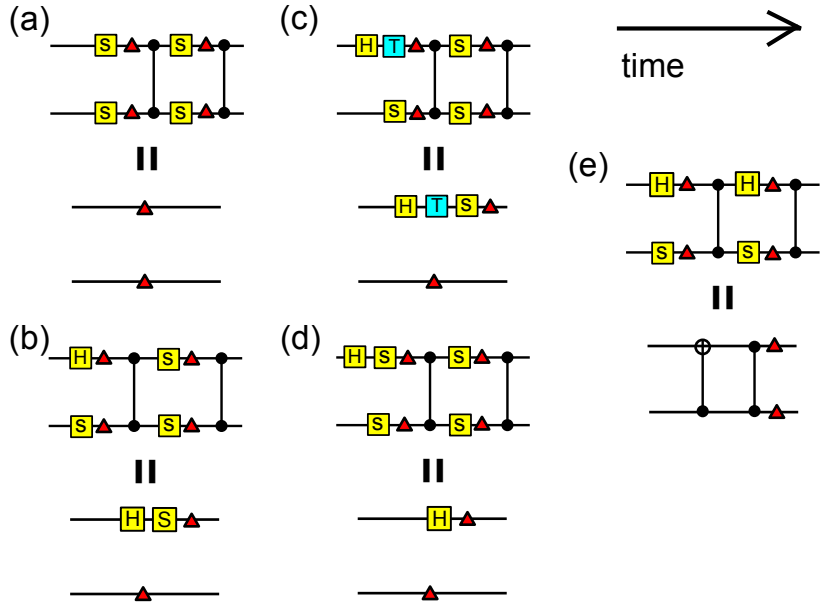}
\end{center}
\caption{(Color online.) 
Operations which can be implemented in the unit cell.
Red triangles are Pauli byproducts.
In (c),  
the blue $T$ can be 
replaced with $T^\dagger$.
} 
\label{unitcell2}
\end{figure}

\begin{figure}[htbp]
\begin{center}
\includegraphics[width=0.3\textwidth]{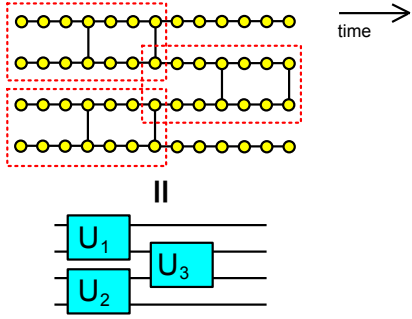}
\end{center}
\caption{(Color online.) 
Tiling for Protocol 2.
} 
\label{tiling}
\end{figure}

%--------------------
\section{Blindness of Protocol 2}

{\bf Theorem 3}:
{\it
Protocol 2 satisfies (B1).
}

{\bf Proof}:
Let $B$ be the random variable which represents the
type of the POVM which Bob performs on his system,
and $M_B$ be the random variable which represents 
the result of the POVM. 
Let $T$ be the random variable which represents Alice's
message to Bob about the channel loss.
Let $A$ be the random variable which represents
Alice's measurement angles.
Bob's knowledge about Alice's measurement angles is given by the conditional probability distribution
of $A=a$ given
$B=b$, $M_B=m_B$ and  $T=t$: 
\begin{eqnarray*}
P(A=a~|~B=b,M_B=m_B,T=t).
\end{eqnarray*}
From Bayes' theorem, we have
\begin{eqnarray*}
P(A=a~|~B=b,M_B=m_B,T=t)
&=&
\frac{P(M_B=m_B,A=a,B=b,T=t)
}
{P(B=b,M_B=m_B,T=t)}\\
&=&
\frac{P(M_B=m_B,A=a,B=b)P(T=t)
}
{P(B=b,M_B=m_B,T=t)}\\
&=&
\frac{P(M_B=m_B~|~A=a,B=b)P(A=a,B=b)P(T=t)
}
{P(B=b,M_B=m_B,T=t)}\\
&=&
\frac{P(M_B=m_B~|~A=a',B=b)P(A=a',B=b)P(T=t)
}
{P(B=b,M_B=m_B,T=t)}\\
&=&
P(A=a'~|~B=b,M_B=m_B,T=t).
\end{eqnarray*}
$\blacksquare$

{\bf Theorem 4}:
{\it
Protocol 2 satisfies (B2).
}

{\bf Proof}:
Let $B$ be the random variable which represents the
type of the POVM which Bob performs on his system,
and $M_B$ be the random variable which represents the
the result of the POVM. 
Let $T$ be the random variable which represents Alice's
message to Bob about the channel loss.
Let $O$ be the random variable which represents
the output of Alice's algorithm.
Bob's knowledge about the output of Alice's algorithm
is given by the conditional probability distribution
of $O=o$ given $B=b$,
$M_B=m_B$, and  $T=t$: 
\begin{eqnarray*}
P(O=o~|~B=b,M_B=m_B,T=t).
\end{eqnarray*}
From Bayes' theorem, we have
\begin{eqnarray*}
P(O=o~|~B=b,M_B=m_B,T=t)
&=&
\frac{P(M_B=m_B,O=o,B=b,T=t)
}
{P(B=b,M_B=m_B,T=t)}\\
&=&
\frac{P(M_B=m_B,O=o,B=b)P(T=t)
}
{P(B=b,M_B=m_B,T=t)}\\
&=&
\frac{P(M_B=m_B~|~O=o,B=b)P(O=o,B=b)P(T=t)
}
{P(B=b,M_B=m_B,T=t)}\\
&=&
\frac{P(M_B=m_B~|~O=o',B=b)P(O=o',B=b)P(T=t)
}
{P(B=b,M_B=m_B,T=t)}\\
&=&
P(O=o'
~|~B=b,M_B=m_B,T=t
).
\end{eqnarray*}
$\blacksquare$

\end{widetext}

%----------------------


\begin{thebibliography}{00}
\bibitem{hammer}
It is insightful to point out here that
one great advantage of the measurement-based quantum computation
which enables a simple blind protocol is the clear separation
between the quantum stage (resource preparation) and the 
classical stage (measurements and classical-signal feedforward).
In order to see it, let us consider the following story.
Alice and Bob are in the same laboratory.
Bob prepares the resource state, and he measures a particle.
Then Alice hammers Bob, and Bob loses his memory.
Bob does the second measurement after waking up
and again Alice hammers Bob.
If they repeat this process until the end of the computation, 
Alice finally gets the output
of the desired quantum computation. However, Bob gets no information
about this quantum computation since he loses his memory every time.
\bibitem{teleportation}
It is interesting
to note that measurement-based quantum computation
is the sequence of teleportations~\cite{Verstraete},
and therefore 
if Alice
does not send her measurement result,
Bob's state is the completely-mixed state.
\bibitem{noQC}
The security of Protocol 1 is easily understood from
the following logic.
In Protocol 1, Alice can choose two strategies:
(1) Doing the correct measurements.
(2) Just discarding all particles sent from Bob.
Let $\rho_i$ $(i=1,2)$
be Bob's state after Alice finishing the strategy $(i)$. 
Of course, $\rho_1=\rho_2$,
since otherwise
Alice can transmit some message to Bob.
And, obviously, $\rho_2$ does not contain
any information about Alice's input, output and algorithm,
since no quantum computation was performed.
Therefore $\rho_1$ does not contain any information about
Alice's computation.
\end{thebibliography}
\end{document}